\documentclass[prl,floatfix,twocolumn,superscriptaddress,showpacs,preprintnumbers,longbibliography,nofootinbib]{revtex4-1}
\usepackage{dcolumn}    
\usepackage{bm} 
\usepackage{graphicx}
\usepackage{amsmath}    
\usepackage{enumitem}
\usepackage{latexsym}
\usepackage{amsfonts}   
\usepackage{amssymb}
\usepackage{array}      
\usepackage{epsfig}
\usepackage[dvipsnames]{xcolor}

\usepackage{color}
\usepackage[colorlinks=true,linkcolor=blue,urlcolor=blue,citecolor=blue,pdfusetitle]{hyperref}
\usepackage{hyperref}
\usepackage{physics}
\usepackage{bbold}

\newcommand{\eq}{Eq.~}
\newcommand{\fig}{Fig.~}

\begin{document}

\title{Witnessing non-classicality through large deviations in quantum optics}
       
\author{Dario Cilluffo}
\email{Corresponding author:dario.cilluffo@unipa.it}
\affiliation{Universit$\grave{a}$  degli Studi di Palermo, Dipartimento di Fisica e Chimica - Emilio Segr\`e, via Archirafi 36, I-90123 Palermo, Italy}
\affiliation{NEST, Istituto Nanoscienze-CNR, Piazza S. Silvestro 12, 56127 Pisa, Italy}
\affiliation{Institut f\"ur Theoretische Physik, Universit\"at T\"ubingen, Auf der Morgenstelle 14, 72076 T\"ubingen, Germany}
\author{Giuseppe Buonaiuto}
 \thanks{The first two authors contributed equally}
\affiliation{School of Physics and Astronomy, University of Nottingham, Nottingham, NG7 2RD, United Kingdom}
\affiliation{Centre for the Mathematics and Theoretical Physics of Quantum Non-equilibrium Systems, University of Nottingham, Nottingham, NG7 2RD, United Kingdom}
\affiliation{Institut f\"ur Theoretische Physik, Universit\"at T\"ubingen, Auf der Morgenstelle 14, 72076 T\"ubingen, Germany}
\author{Salvatore Lorenzo}
\affiliation{Universit$\grave{a}$  degli Studi di Palermo, Dipartimento di Fisica e Chimica - Emilio Segr\`e, via Archirafi 36, I-90123 Palermo, Italy}
\author{G. Massimo Palma}
\affiliation{Universit$\grave{a}$  degli Studi di Palermo, Dipartimento di Fisica e Chimica - Emilio Segr\`e, via Archirafi 36, I-90123 Palermo, Italy}
\affiliation{NEST, Istituto Nanoscienze-CNR, Piazza S. Silvestro 12, 56127 Pisa, Italy}
\author{Francesco Ciccarello}
\affiliation{Universit$\grave{a}$  degli Studi di Palermo, Dipartimento di Fisica e Chimica - Emilio Segr\`e, via Archirafi 36, I-90123 Palermo, Italy}
\affiliation{NEST, Istituto Nanoscienze-CNR, Piazza S. Silvestro 12, 56127 Pisa, Italy}
\author{Federico Carollo}
\affiliation{School of Physics and Astronomy, University of Nottingham, Nottingham, NG7 2RD, United Kingdom}
\affiliation{Centre for the Mathematics and Theoretical Physics of Quantum Non-equilibrium Systems, University of Nottingham, Nottingham, NG7 2RD, United Kingdom}
\affiliation{Institut f\"ur Theoretische Physik, Universit\"at T\"ubingen, Auf der Morgenstelle 14, 72076 T\"ubingen, Germany}
\author{Igor Lesanovsky}
\affiliation{School of Physics and Astronomy, University of Nottingham, Nottingham, NG7 2RD, United Kingdom}
\affiliation{Centre for the Mathematics and Theoretical Physics of Quantum Non-equilibrium Systems, University of Nottingham, Nottingham, NG7 2RD, United Kingdom}
\affiliation{Institut f\"ur Theoretische Physik, Universit\"at T\"ubingen, Auf der Morgenstelle 14, 72076 T\"ubingen, Germany}

\begin{abstract}
Non-classical correlations in quantum optics as resources for quantum computation are important in the quest for highly-specialized quantum devices. Here, we put forward a methodology to witness non-classicality of the output field from a generic quantum optical setup via the statistics of time-integrated photo currents. Specifically, exploiting the thermodynamics of quantum trajectores, we express a known non-classicality witness for bosonic fields fully in terms of the source master equation, thus bypassing the explicit calculation of the output light state.
\end{abstract}
\date{\today}
\maketitle

\begin{figure}[htbp]
\includegraphics[scale=1.4,angle=0]{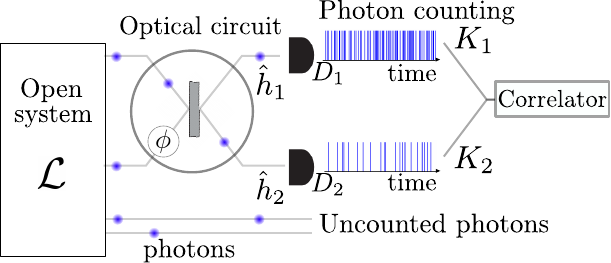}
\caption{\textbf{Sketch of dissipative quantum optical network.}
A generic quantum network is composed of a series of interconnected elements (system) emitting continuously detected radiation in the environment. The emission fields can be manipulated and transformed via a series of unitary operations in an optical circuit using beam splitters and phase shifters. The photo-detectors ($D_1$ and $D_2$) allows the simultaneous reconstruction of the quantum trajectories for two emission channels. Their correlation properties are studied to uncover quantum non-classicality of radiation.}
\label{scheme}
\end{figure}

\textit{Introduction.}
During the last decades, several platforms have been proposed for implementing efficiently quantum computing \cite{FeynmanIJP1982,LossPRA1998,BlaisPRA2004}: all of them suffer from the effect of decoherence, given by the coupling to the environment \cite{preskill1998lecture}, which ultimately deteriorate the non-classical properties of the systems considered. In fact, for a quantum computational scheme to outperform a classical one, one requires that at least one of its component exhibits genuinely quantum features \cite{MariPRL2012}. 
When the environment is the electromagnetic vacuum causing photon emission, such as in dissipative optical networks \cite{PauleBook2018}, the statistical analysis of the output light contains the information about the dynamical features of the open quantum systems \cite{carmichaelbook2009}. In particular, the emitted photons can be used as a resource for quantum information processing \cite{KokRMP2007}. Hence, the detection and optimization of non-classical correlations in the photons emitted by a general optical setup is of primary relevance for a  variety of technological applications.
In this work, we present a methodology to witness non-classicality of the light emitted from a generic quantum optical setup via the statistics of time-integrated photo currents.
 Specifically, the type of setups we consider includes an open quantum system, which is the source of photons, and an optical circuit used to manipulate the emission, as shown in \fig\ref{scheme}. To obtain the statistical properties of the photons arriving at the detectors we make use of the \textit{large deviations} approach \cite{ellis1996overview,touchette2009large,vulpiani2014large,manzano2014symmetry}. This allows to access to the joint probability distribution of the photon counting at long times, together with relevant statistical quantities such as the fluctuations of the counting fields and corresponding cross-correlation functions.
In this way a non-classicality criterion is formulated based on the time-integrated observables of the detection \cite{garrahan2010thermodynamics,garrahan2011quantum,BuonaiutoNJOP2019,CilluffoJSTAT2019}.
Theoretically, this establishes, from the theoretical point of view, a natural link between the statistical-physics approach for analyzing the output and the dynamics of open quantum systems \cite{garrahan2010thermodynamics}, and a general class of non-classicality measures in quantum optics.
We provide simple but instructive examples, where non-classical correlations are witnessed in different dynamical regimes of the sources, and for a broad range of parameters of the components of the optical circuit. 
Our theoretical scheme is effective in predicting the outcomes of quantum optics experiments that make use of photon countings to witness non-classicality \cite{SilberhornPRL2015,HarderPRL2016,BarbieriPRL2016,HarderPRA2014,AvenhausPRL2010,PevrinaPRA2017}. 

\textit{Open quantum systems and Large Deviation.}
Our goal is to access the statistical properties of the output light of an open quantum system emitting into $N_L$ different modes called $\mathcal{B}_\mu$, with $\mu=1,...,N_L$.
The photon counting statistics at the detectors (see \fig\ref{scheme}) provides information about the state of the open system as well as about the features of the optical circuit \cite{carmichaelbook2009}. The counting statistics is fully characterized by the cumulants of the associated photon-counting probability distribution, which are encoded in the \textit{scaled cumulant generating function} (SCGF). Next we briefly review how to compute the SCGF in a rather general setting.
The evolution of the reduced density operator of the open system $\rho$, in the Markovian approximation, is given by the well-known Lindblad master equation \cite{haroche2006exploring,lindblad1976generators,gorini1976completely}, 
\begin{equation}
 \dot{\rho} = -i [\hat{H}, \rho] + \sum_{\mu=1}^{N_L} \mathcal{D}(\hat{L}_\mu)\rho \equiv \mathcal{L}[\rho],
\end{equation} 
 where the \textit{jump operator} $\hat{L}_\mu$ coresponds to the interaction with the field mode $\mathcal{B}_\mu$ and $\mathcal{D}(\hat{L}_\mu)[\rho] =\sum_{\mu=1}^{N_L} [ \hat{L}_\mu \rho \hat{L}^\dagger_\mu - \tfrac{1}{2} \{\hat{L}_\mu^\dagger \hat{L}_\mu, \rho\}]$. Following a standard approach of open quantum system theory \cite{haroche2006exploring}, we gather information about the evolution of $\rho$ by continuous monitoring of the environment.

Let us divide our jump operators in $N$ subsets, $\mathcal{J}_i$, each of size $n_i$, with $i=1,...,N$, and $\sum_{i=1}^{N} n_i = N_L$: suppose we record the occurrence of jump events due to the action of the operators in the first $M$ subsets ($M < N$), and let $K_m$ be the absolute number of detected jumps in time (\textit{counting field}) corresponding to each subset $\mathcal{J}_m$ with $m =1,2...,M$. 
Furthermore we assume that the action of these jump operators induces photoemission.
In short notation, we define the vector $\mathbf{K}=(K_1, K_2,...,K_M)$ to be the photon countings associated with each $\mathcal{J}_m$.
The probability to observe $\mathbf{K}$ counts from each decay channel after a time $t$ is $ P_t(\mathbf{K}) = \operatorname{Tr}\{\rho^{\mathbf{K}}(t)\}$, where $\rho^{\mathbf{K}}(t)$ is the un-normalized reduced density operator conditioned to $\mathbf{K}$ \cite{ZollerPRA1987}.
The moment generating function associated with $ P_t(\mathbf{K})$ reads
$Z_t(\mathbf{s}) = \sum_{K=0}^{\infty} P_t(\mathbf{K}) e^{-\mathbf{s} \cdot \mathbf{K}}$ with $\mathbf{s}=(s_1,...,s_M)$. Here $s_m$ is the \textit{conjugated field} corresponding to $K_m$. 

The outcomes of photocount experiments are time-integrated photocurrents 
\begin{equation}
\langle k_i \rangle = \frac{1}{t} \sum_{j=1}^{n_i} \operatorname{Tr}\left\{\int_0 ^t d\tau \hat{L}_j^\dagger \hat{L}_j \rho(\tau)\right\}
\label{current}
\end{equation}
with $i = 1, 2,...,M$. For $t$ much greater than the typical timescale of the system $\tau_c$, the probability distribution associated to the photon counting measurement takes a large deviation form \cite{touchette2009large}. Specifically, at long times the moment generating function can be asymptotically approximated through the large deviation theory as an exponential function of time as
\begin{align}
Z_t(\mathbf{s}) \sim e^{t \theta(\mathbf{s})}.
\label{fundamental}
\end{align}
This basically expresses the large deviation principle for the moment generating function. The analogue for the count probability reads $P_t(\mathbf{K}) \sim e^{t \varphi(\mathbf{K}/t)}$, where $\varphi(x) = - \min_{s} \{ x s + \theta(s)\}$.
The function $\theta(\mathbf{s})=\tfrac{1}{t} \ln Z_t (\mathbf{s})$ is the SCGF. It can be proven \cite{touchette2009large,garaspects} that this is given by the maximum real eigenvalue of the deformed superoperator
\begin{align}
\mathcal{L}_s[\rho] = \mathcal{L} [\rho] -\sum_{i=1}^{M}(1-e^{-s_i})\sum_{{\mu_i}=0}^{n_i} \hat{L}_{\mu_i} \rho \hat{L}^\dagger_{\mu_i} ,
\label{tilted}
\end{align}
which features the standard Liouvillian and the dissipator, with the jump parts corresponding to each subset $\mathcal{J}_i$, the latter being weighted by the factor $e^{-s_i}$.
The cumulants of the distribution $P_t(\mathbf{K})$ at long times are given by the derivative of $\theta(\mathbf{s})$ at $\mathbf{s}=0$: cumulants give direct access to the moments of the associated distribution \cite{JordanBook1965}.

In this work, for the sake of clarity, we consider the case $M=2$ and $n_1=n_2=1$, i.e., two distinct counting fields each associated with a single jump operator, as shown in \fig\ref{scheme}.
Then \eq\eqref{tilted} takes the form $\mathcal{L}_{s_1,s_2}[\rho] = \mathcal{L} [\rho] - \sum_{\mu=1}^{2} (1-e^{-s_\mu}) \hat{L}_\mu \rho \hat{L}^\dagger_\mu,$
and the maximum real eigenvalue of $\mathcal{L}_{s_1,s_2}$ is $\theta(s_1,s_2)=\tfrac{1}{t} \ln {Z}_t (s_1, s_2)$, with ${Z}_t (s_1, s_2)$ the moment generating function of the probability distribution ${P}_t(K_1,K_2)$ associated with the photocount measurement described by the jump operators $\hat{L}_\mu$ in the long-time limit.
In particular, we recover the moments of the marginal distributions $P(K_1)$ and $P(K_2)$ by setting $s_1 = 0$ or $s_2 = 0$.
By exploiting the double weighting it is possible to access to the correlations between the counting fields at the detectors. In particular the covariance reads
\begin{equation}
 \operatorname{cov}(k_1, k_2) = \langle k_1 k_2 \rangle - \langle k_1 \rangle \langle k_2 \rangle = \partial_{s_1} \partial_{s_2} \theta(s_1, s_2) |_{s_1=s_2=0}.
\end{equation}
All other moments can be easily recovered in terms of higher order derivatives of $\theta(s_1,s_2)$.
The possibility of accessing the full statistics of the joint probability distribution, as we shall see in the following, allows to make use of non-classicality measures on the bath operators, with the idea of finding possible signatures of quantum correlations between the detection events (in the long-time limit).
\begin{figure*}[!htbp]
\includegraphics[scale=0.91,angle=0]{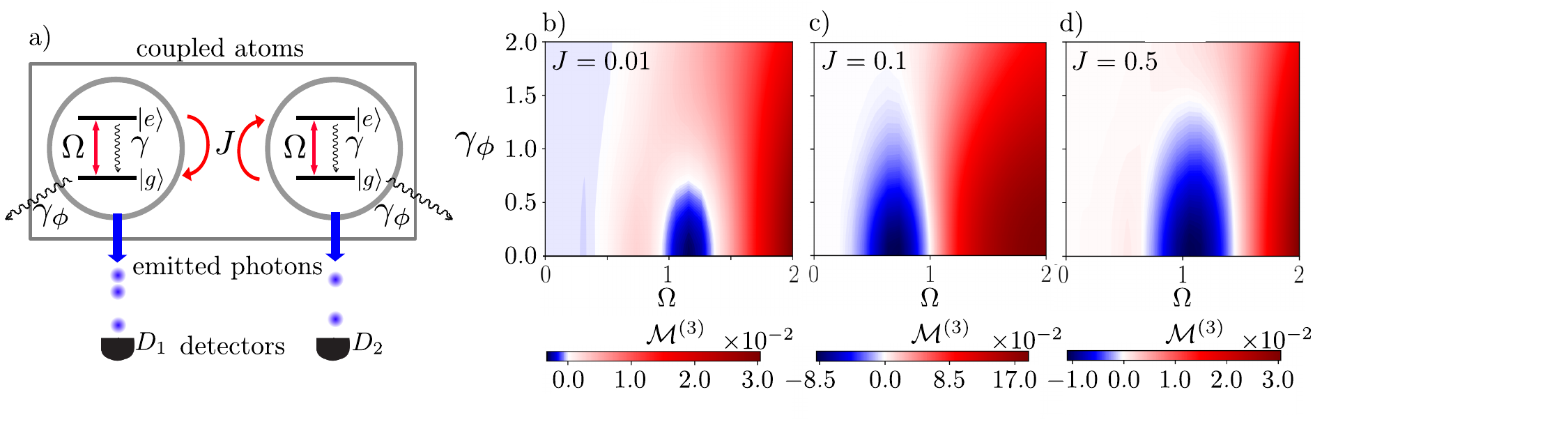}
\caption{\textbf{Non-classicality witness for emission from coupled atoms}. 
(a) third-order Vogel's determinant for a system of two coherently-driven interacting atoms (coupling strength $J$) subject to dephasing, as a function of the dephasing rate $\gamma_\phi$ and Rabi frequency $\Omega$. Plots (b)-(d) are for different coupling strengths: $J=0.01$ (b), $J=0.1$ (c), $J=0.5$ (d). In all cases we observe a sharp separation between classical (positive Vogel's determinant) and quantum states of the emitted radiation (negative regions).
}
\label{vogelfigdeph}
\end{figure*}

\textit{Vogel's non-classicality criterion (VC).} 
This criterion \cite{SchuckinPRA2005,Shchukin_rapid} gives a necessary and sufficient condition to establish whether correlations in a stationary radiation field are nonclassical or not. It consists of a rephrasing of the well-known non-classicality criterion based on the negativity of the Glauber-Sudarshan distribution (or \textit{$\mathcal{P}$-distribution})  \cite{GlauberPR1963,CahillPR1969} in terms of photon-counting detection.
Referring to the setup in \fig\ref{scheme}, let us consider the generic bosonic operators $\hat{h}_{i}$, $(i=1,2)$, of the two output fields, and assume they are normally-ordered functions of the associated destruction and creation operators $\hat{a}_i$ and $\hat{a}_i^\dagger$ of the each mode. A generic operator acting on the two-mode field is defined as $\hat{f} = \sum_{n, m =0} ^{\infty} f_{n m} \hat{h}_1^{\dagger n} \hat{h}_2^m$, which is a normally-ordered power series of $\hat{h}_i $ and $ \hat{h}_i^\dagger$. 
The expectation value of $\langle : \hat{f}^\dagger \hat{f} : \rangle$ reads:
\begin{align}
\langle : \hat{f}^\dagger \hat{f} : \rangle &= 
\sum_{n, m, k, l =0} ^{\infty} f_{n m} f_{k l}^*  \langle \hat{h}^{\dagger n+k}_1 \hat{h}^{m+l}_2 \rangle \nonumber
\\
&= \int_{\mathbb{C}} \mathcal{P}(\alpha_1,\alpha_2) |f(\alpha_1,\alpha_2)|^2 d^2 \alpha_1 d^2 \alpha_2 ,
\label{1mode}
\end{align}
where the last equality follows from the optical equivalence theorem \cite{Schleich2011}, $f(\alpha_1,\alpha_2) = \sum_{n, m =0} ^{\infty} f_{n m} \hat{h}^{\dagger n}_{1}(\alpha_1,\alpha_1^*) \hat{h}^m_{2}(\alpha_2,\alpha_2^*)$ and where $\mathcal{P}(\alpha_1,\alpha_2)$ is the Glauber-Sudarshan distribution.
Since $\langle : \hat{f}^\dagger \hat{f} : \rangle < 0$ entails $\mathcal{P}(\alpha_1,\alpha_2) < 0$ for some points $(\alpha_1,\alpha_2)$ of the phase-space, the negativity of \eq\eqref{1mode} is a clear signature of non-classicality in radiation fields. Note that \eq\eqref{1mode} is a quadratic form and is non-negative iff all the principal minors of matrix $\mathcal{M}_{nm,kl}= \langle \hat{h}_1^{n + k}  \hat{h}_2^{m+l}\rangle$ (see \eq\eqref{mommatr} in Appendix)
are positive, according to the Sylvester criterion \cite{SperlingPRA2013}.
Referring to the setup in \fig\ref{scheme} and according to \cite{SperlingPRA2013,SchuckinPRA2005}, we express the VC in terms of click-counting operators, which, from the open quantum system point of view, take the form $\hat{h}_j=\hat{L}_j^\dagger \hat{L}_j$. 
Thus the elements of $\mathcal{M}_{nm,kl}$ are the moments of the photon-counting stationary distribution $P(K_1,K_2)$, which gives the probability to record $K_1$ clicks at photodetector $D_1$ and $K_2$ at $D_2$. Hence, the criterion is now formulated in terms
of time-integrated functions, like the photocurrents defined in \eq\eqref{current}.
The moments in $\mathcal{M}_{nm,kl}$ are easily calculated through iterative derivation of two-mode moment generating function associated to 
 $P(K_1,K_2)$.
 Note that the mixed derivatives of the double-biased scaled cumulant generating function $\theta(s_1,s_2)$ give us the mixed scaled cumulants directly linked to the two-mode moments in $\mathcal{M}_{nm,kl}$.

Different setups have been proposed, realized and successfully used \cite{HarderPRA2014,AvenhausPRL2010,PevrinaPRA2017} in order to measure the click-counting distribution thus uncovering quantum correlations of radiation fields.
The click-counting distribution can approximate $P(K_1,K_2)$ involving photon-counting via a long-time measurement through photon-number-resolving detectors. As shown in \cite{SperlingPRA2013}
 once the estimate of the stationary probabilities are known it is clearly possible to recover the moments in $\mathcal{M}_{nm,kl}$. Usually, the higher the order of the moment we calculate, the less accurate our estimate will be. In the cases we study next, low-order moments are enough to determine non-classical features of radiation. 
It was shown \cite{SperlingPRA2013} that the binomial form for the click-counting probability distribution holds for any positive-operator valued measurement (POVM) either linear or non-linear in the number of emitted photons. Thus the large deviation formalism allows us to inherently access all the cumulants associated to any photon-counting process defined by the unraveling of the master equation.

\textit{Non-classicality in dissipative circuits.}
Typical coherent and squeezed radiation sources (pumped cavities, nonlinear active media) can be studied from the point of view of open quantum system theory \cite{carmichaelbook2009}.
Referring to the generic setup in \fig\ref{scheme}, we now consider two different source structures: 
a pair of coupled two-level atoms, each coherently driven and subject to decay in its own emission channel and two non-interacting atoms whose outputs are correlated via a beam splitter and a phase shifter. In both cases we introduce dephasing on each atom with rate $\gamma_{\phi}$: such dephasing channel spoils coherence, hence it is expected to affect non-classicality of emitted light.
\\
\textit{Two coupled atoms.} The total Hamiltonian of the system reads 
\begin{align}
\hat{H}=&\sum_{i=1}^{2} [\tfrac{\Omega}{2} ( \hat{\sigma}^+_i + \hat{\sigma}^-_i) + \sqrt{\gamma}\left( \hat{\sigma}^+_i \hat{a}_i+ H.c.\right)]\nonumber
\\&+ J \left( \hat{\sigma}^+_1 \hat{\sigma}^-_2 + H.c.\right),
\end{align}
where $\gamma$ is the decay rate of the each atom, $\Omega$ the Rabi frequency, $\hat{\sigma}^+_i$ and $\hat{\sigma}^-_i$ are the ladder operators, $\hat{a}_i$ is the annihilation operator of the bosonic mode coupled to the $i$th atom\footnote{$\hat{a}_i$ operators are intended as time-mode bosonic operator, or input modes i.e. Fourier transform of field normal mode operators $\hat{a}_\omega$ under the assumption of white coupling between system and environment \cite{gardiner2004quantum}.}
 and $J$ is the coupling strength.
\begin{figure}
\includegraphics[scale=0.93,angle=0]{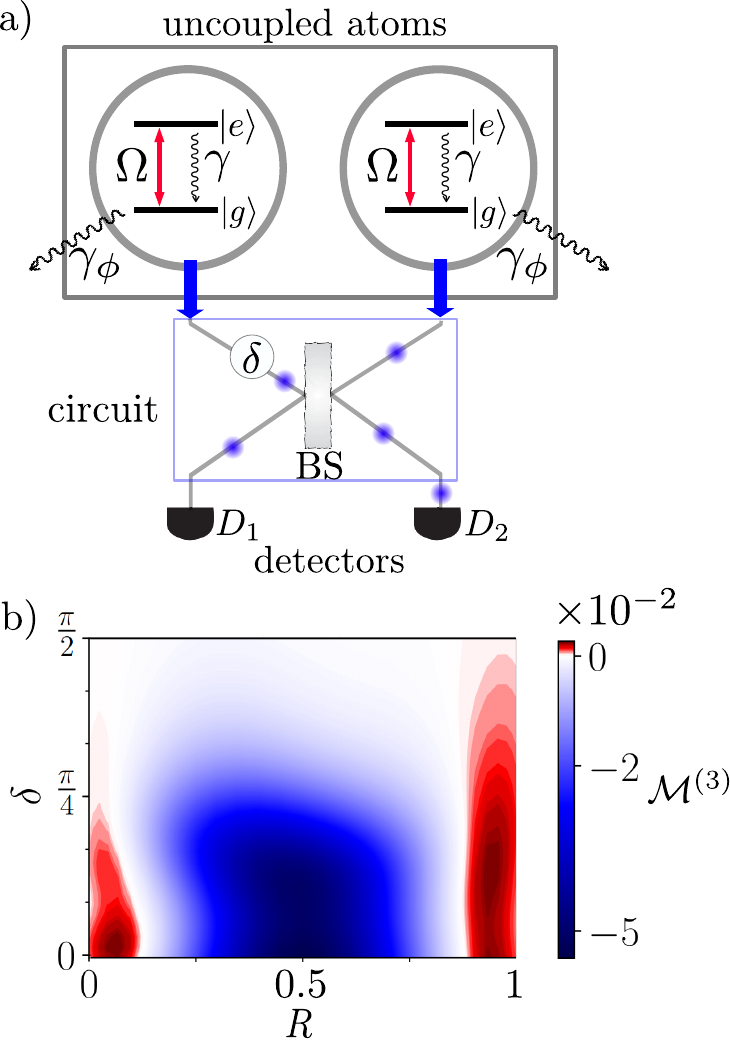}
\caption{\textbf{Non-classicality witness for emission from optical circuit}. 
The system (a) is composed of a two coherently-driven non-interacting atoms subject to dephasing $\gamma_\phi$, emitting into the input channel of a generic unitary circuit composed of a phase-shifter and a beam splitter.
(b) We show the value of the third-order Vogel determinant as a function of reflectivity $R=\sin^2 \zeta$ and phase-shift $\delta$.
}
\label{vogelfig2}
\end{figure}
The jump operators of this elementary network are thus $\hat{J}_1= \sqrt{\gamma} \hat{\sigma}^-_1$ and  $\hat{J}_2= \sqrt{\gamma} \hat{\sigma}^-_2$. 
We can straightforwardly compute the large deviation moments matrix and the corresponding Vogel determinants for the joint photon counting probability distribution. It is worth noting that the second-order principal minor ($\mathcal{M}^{(2)}$) does not contain information on the cross-correlations between the emitted field, which is our focus. Thus, it is necessary to consider the next order minor.
A numerical investigation of the third-order principal minor ($\mathcal{M}^{(3)}$, see Appendix) reveals the presence of quantum correlations between detection events in the emission channels. \fig\ref{vogelfigdeph} shows $\mathcal{M}^{(3)}$ as a function of the Rabi frequency $\Omega$ and dephasing rate $\gamma_{\phi}$ for three values of the coupling rate $J$. In each case, non-classicality is reduced as the dephasing rate grows. Negativity grows with $\Omega$, reaching a maximum and then saturating to a positive value. Dephasing destroys quantum coherences making the atoms behave like classical objects, and this results in classical radiation fields, as expected. Higher values of $\Omega$ speed up Rabi oscillations: the effective coarse-graining time-integration is lower bounded by $1/\gamma$. Hence, we expect the time integrated photo-current becomes insensitive to the intensity fluctuations, resulting in a crossover between negative and non-negative values of the determinant. 
Furthermore we notice that the absolute minimum of the third-order determinant does not grow linearly with the coupling strength, but rather decreases when increasing $J$ . It is indeed expected that the strong coupling between the two atoms makes the emission less likely to happen \cite{BambaPRA2011}. The strong coupling contribution results in an effective shift of the energy level of the system and the perfect resonance condition is lost: the dominant component of the output fields becomes vacuum, hence reducing the amount of cross correlations.

\textit{Non-interacting atoms and unitary circuit.} 
We consider next the case in which correlations can arise by \textit{processing} the emitted fields of two non-interacting atoms ($J=0$) through a unitary transformation employing a beam splitter ($\hat{\mathcal{U}}_{BS} = \cos\zeta ~ \mathbb{1} + i \sin\zeta ~ \hat{\sigma}_x$) and a phase shifter (\fig\ref{vogelfig2}).
The corresponding jump operators read $\hat{J}_1= \sqrt{\gamma_1} \cos \zeta ~ \hat{\sigma}^-_1 + i  \sqrt{\gamma_2} \sin \zeta ~ \hat{\sigma}^-_2$ and  $\hat{J}_2=  i \sqrt{\gamma_1} \sin \zeta ~ \hat{\sigma}^-_1 +   \sqrt{\gamma_2} \cos \zeta ~ \hat{\sigma}^-_2$. 
We set $\Omega =0.5 \gamma$ and $\gamma_\phi =0.1$ and study non-classicality as a function of the reflectivity $R=\sin^2 \zeta$ and the phase difference $\delta$ between the two channels due to the phase shifter. For total transmission ($\zeta=0$) and total reflection ($\zeta=\pi/2$), we notice that the determinant is positive. The maximum negativity is reached for a $50/50$ beam splitter and decreases as the phase-shift $\delta$ grows. Thus, by adjusting appropriately the parameters of the optical circuit, such as the relative phase shift $\delta$, it is possible to enhance or destroy quantum interference effects of the output state.

\textit{Conclusions.} 
In summary, we have shown how to detect signatures of non-classicality through the statistics of time-integrated quantities, such as the photon counts. This offers the possibility to benchmark approaches for producing quantum resources for information and computation via general optical circuits and open quantum systems.
Our findings can be extended both to inperfect detection as well as to recently proposed high-performing photon-number-resolving detection schemes \cite{malz2019number}. Finally, we point out here a possible outlook of this work: the formalism here developed can be implemented to tackle the problem of characterizing many-body phases of matter by analysing the statistical properties of emitted and scattered photons or bath quanta in general.

\textit{Acknowledgements.}
We thank for fruitful discussions Fabio Sciarrino, Taira Giordani, Fulvio Flamini, Iris Agresti and Alessia Castellini.
D. C. acknowledges the University of Nottingham for the hospitality.
The research leading to these results has received funding from the European Union's H2020 research and innovation programme [Grant Agreement No. 800942 (ErBeStA)]. 
We acknowledge support under PRIN project 2017SRN-BRK QUSHIP funded by MIUR.

Dario Cilluffo and Giuseppe Buonaiuto contributed equally to the realization of the work.

\bibliographystyle{apsrev4-1}
\bibliography{biblio}

\clearpage
\section{Appendix}
In line with \cite{SchuckinPRA2005} 
we put the elements $\mathcal{M}_{nm,kl}$ in ascending order 
with respect of the sum of the couples of indexes $(n+m)$ and such that $(n,m) < (n - 1 , m + 1)$. The resulting matrix reads:
\begin{align}
 \mathcal{M} = 
 \left( \begin{matrix}
  1 & \langle \hat{h}_1 \rangle & \langle \hat{h}_2 \rangle & \langle \hat{h}^2_1 \rangle & \langle \hat{h}_1 \hat{h}_2 \rangle &\dots\\
  \langle \hat{h}_1 \rangle & \langle \hat{h}_1^2 \rangle & \langle \hat{h}_1 \hat{h}_2 \rangle&
  \langle \hat{h}_1^3 \rangle &
  \langle \hat{h}_1^2 \hat{h}_2 \rangle &\dots\\
  \langle \hat{h}_2 \rangle &
  \langle \hat{h}_1 \hat{h}_2 \rangle&
  \langle \hat{h}_2^2 \rangle&
  \langle \hat{h}_1^2 \hat{h}_2 \rangle&
  \langle \hat{h}_1 \hat{h}_2^2 \rangle &\dots\\
  \vdots & \vdots & \vdots & \vdots & \vdots  
 \end{matrix}\right) ,
\label{mommatr}
 \end{align}
where each element 
is a moment of the bivariate counting probability distribution associated to the counting operators. 
In the case of Large Deviation calculation our average includes an integration over the duration time of trajectories (as in \eq\eqref{current}), thus we have direct access to the scaled time-integrated cumulant of such photon-counting probability distribution. The moments resulting by combination of scaled cumulants will be scaled in turn. 
From \eq\eqref{fundamental} we have that $\mathcal{M}_{nm,kl}=\frac{\partial^{n+k}}{\partial s_1^{n+k}} \frac{\partial^{m+l}}{\partial s_2^{m+l}} Z_t(s_1,s_2) \propto t^{n+m+k+l}$, i.e. the power of $t$ depends only on the position of the element in the matrix.
Thus the power function of $t$ multiplying each summand featured in the determinant is invariant under permutation of indexes, the scaling resulting only in an overall positive factor $t^{f(N)}$ multiplying each $N$-th order Vogel determinant.
Up to a multiplicative constant, the third order principal minor we used in the examples reads (s-dependencies are omitted):
\begin{align}
&\mathcal{M}^{(3)} = 2\left(\frac{\partial \theta}{\partial s_1} \frac{\partial \theta}{\partial s_2}\right)\left(\frac{\partial^2 \theta}{\partial s_1 \partial s_2} +\frac{\partial \theta}{\partial s_1} \frac{\partial \theta}{\partial s_2}\right)\nonumber\\&-\left(\frac{\partial^2 \theta}{\partial s_1 \partial s_2} + \frac{\partial \theta}{\partial s_1} \frac{\partial \theta}{\partial s_2} \right)^2 - \left(\frac{\partial \theta}{\partial s_1}\right)^2\left(\frac{\partial^2 \theta}{\partial s_1 \partial s_2} + \left(\frac{\partial \theta}{\partial s_2}\right)^2\right)\nonumber \\&- \left(\frac{\partial \theta}{\partial s_2}\right)^2 \left(\frac{\partial^2 \theta}{\partial^2 s_1} + \left(\frac{\partial \theta}{\partial s_1}\right)^2 \right) \nonumber \\&+ \left(\frac{\partial^2 \theta}{\partial s_1 \partial s_2} + \left(\frac{\partial \theta}{\partial s_2}\right)^2\right) \left(\frac{\partial^2 \theta}{\partial s^2_1} + \left(\frac{\partial \theta}{\partial s_1}\right)^2\right).
\end{align}
evaluated at point $s_1=s_2=0$.
\end{document}